\documentclass[a4paper]{spie}

\usepackage{amsmath,amsfonts,amssymb}
\usepackage{graphicx}
\usepackage[colorlinks=true, allcolors=blue]{hyperref}

\title{Advances in High-Precision Astrometry for Exoplanet Detection for Astrometric Missions}

\author[a]{Fabien Malbet}
\author[b]{J\'er\^ome Amiaux}
\author[b]{Florence Ardellier-Desages}
\author[b]{\'Eric Doumayrou}
\author[b]{Pierre-Antoine Frugier}
\author[b]{Pierre-Olivier Lagage}
\author[c]{Alain L\'eger}
\author[a,d,e]{Manon Lizzana}
\author[b]{Jerôme Martignac}
\author[d]{Julien Michelot}
\author[a]{Fabrice Pancher}
\author[b]{Thibault Pichon}
\author[b]{Samuel Ronayette}
\author[a]{Hugo Rousset}
\author[a]{S\'ebastien Soler}
\author[f]{\'Eric Thi\'ebaut}

\affil[a]{Univ. Grenoble Alpes, CNRS, IPAG, F-38000 Grenoble, France}
\affil[b]{Univ. Paris-Saclay, Univ. Paris Cit\'e, CEA, CNRS, AIM, France}
\affil[e]{Universit\'e Paris-Saclay, CNRS, CNES, IAS, Gif-sur-Yvette, France}
\affil[c]{CNES, F-75001 Paris, France}
\affil[d]{Pyxalis, F-38430 Moirans, France}
\affil[f]{Universit\'e Lyon 1, ENS de Lyon, CNRS, CRAL, UMR 5574, Saint-Genis-Laval, France}

\authorinfo{Corresponding author: Fabien Malbet\\E-mail: Fabien.Malbet@univ-grenoble-alpes.fr}

\begin{document}
\maketitle

\begin{abstract}
High-precision astrometry is a cornerstone for detecting Earth-like exoplanets in the habitable zones of nearby solar-type stars and probing the nature of dark matter in galactic environments. Achieving sub-microarcsecond accuracy requires unprecedented control over detector morphology, optical distortion, and astrometric measurement techniques.

This paper reviews recent progress in characterizing large format detector at the micropixel scale level, calibrating optical distortion using reference stars, and refining differential astrometry methods to detect the faint astrometric signatures of Earth-mass exoplanets.

We highlight the latest developments in the RETINA project, which aims to advance the TRL of GIGAPYX 220 MP CMOS detectors from Pyxalis which are critical components for future space missions such as ESA's Theia and NASA's Habitable Worlds Observatory (HWO). The project focuses on precise calibration methods, including Young's fringe projection, to monitor pixel geometry and intra-pixel response across gigapixel-scale focal planes. By integrating these advancements, RETINA seeks to enable astrometric detection of Earth-mass exoplanets with a precision of few tens of micropixels on the focal plane.

We discuss the technical strategies, laboratory validations, and system-level integration required to achieve these goals, emphasizing the synergy between detector technology, optical calibration, and mission design. These innovations pave the way for next-generation astrometric instruments capable of transforming our understanding of exoplanetary systems and dark matter dynamics.
\end{abstract}


\section{INTRODUCTION}
\label{sec:intro}

\begin{figure}[t]
  \centering
  \includegraphics[width=0.9\hsize]{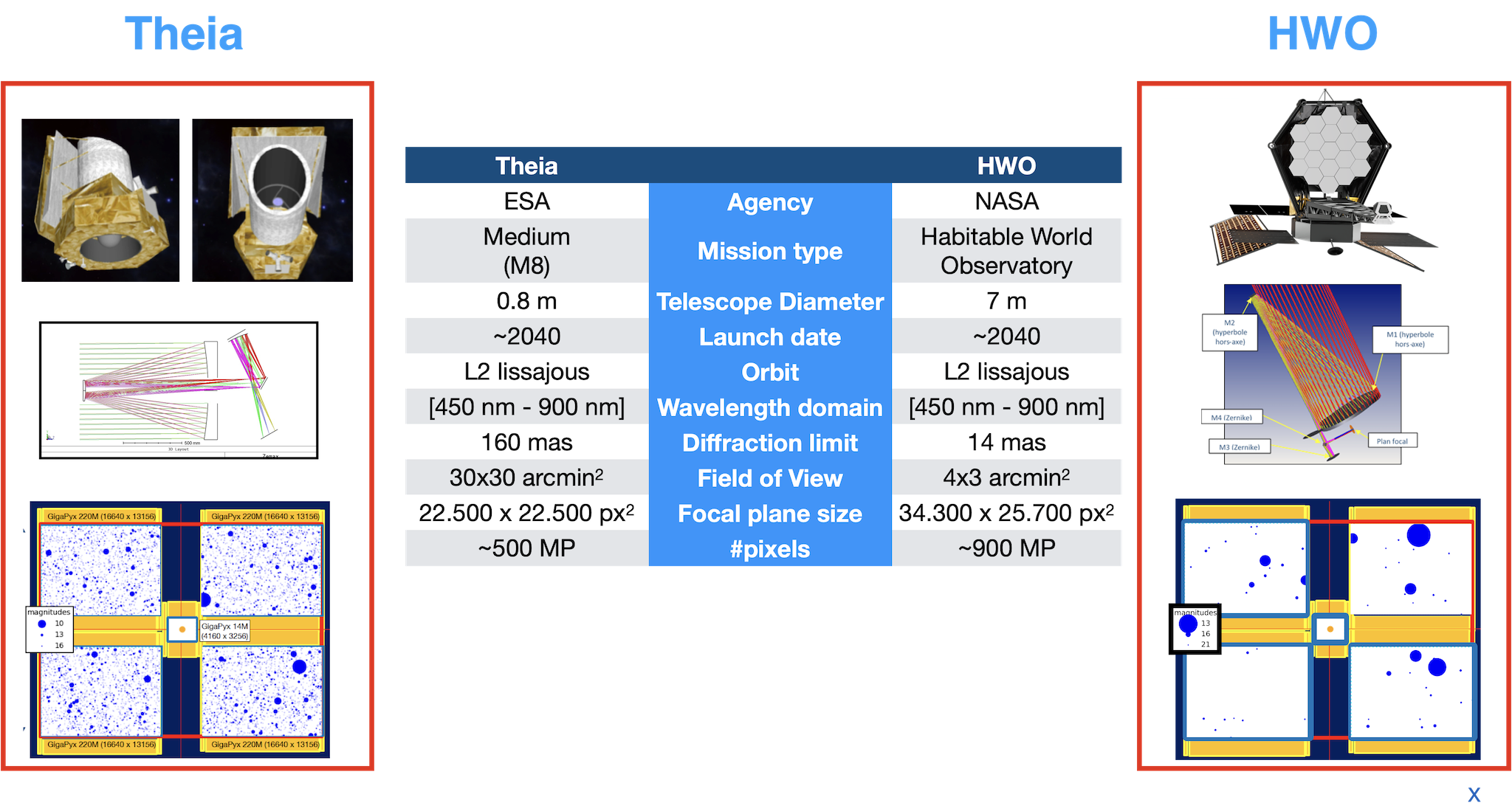}
  \caption{Candidate space missions for high precision astrometry: Theia for an ESA medium mission, and the NASA Habitable Worlds Observatory in astrometric mode.}
  \label{fig:missions}
\end{figure}

High-precision astrometry has emerged as a transformative tool for addressing two of the most compelling questions in modern astrophysics: the detection and characterization of Earth-like exoplanets in the habitable zones of nearby stars, and the exploration of dark matter dynamics in galactic environments\cite{2021ExA....51..845M}. Unlike other exoplanet detection methods\cite{2018haex.bookE.196M}, astrometry provides direct measurements of stellar reflex motion, yielding unambiguous mass estimates and three-dimensional orbital parameters. For Earth-mass planets orbiting Sun-like stars at distances of 10 parsecs, the astrometric signal is on the order of 0.3 microarcseconds, a precision that demands unprecedented control over instrumental systematics.

To meet this challenge, candidate space missions such as ESA's Theia\cite{2021ExA....51..845M} and NASA's Habitable Worlds Observatory (HWO)\cite{2024SPIE13092E..1NF} are being designed with astrometric instruments capable of sub-$\mu$as precision (Fig.~\ref{fig:missions}). These instruments rely on gigapixel-scale focal planes composed of large-format CMOS detectors, whose performance and calibration must be pushed to their limits. This paper synthesizes recent advancements\cite{ArdellierDesages2026_TechMaturation, Amiaux2026_SystemAnalysis, Lizzana2026_Calibrations, Rousset2026_YoungsFringes, Pancher2026_Benches} in this domain, with a particular focus on the RETINA project, a technological maturation effort aimed at enabling the next generation of high-precision astrometric observations.

\section{RETINA: ADVANCING DETECTOR TECHNOLOGY FOR ASTROMETRY}
\label{sec:retina}

\begin{figure}[t]
  \centering
  \includegraphics[width=0.9\hsize]{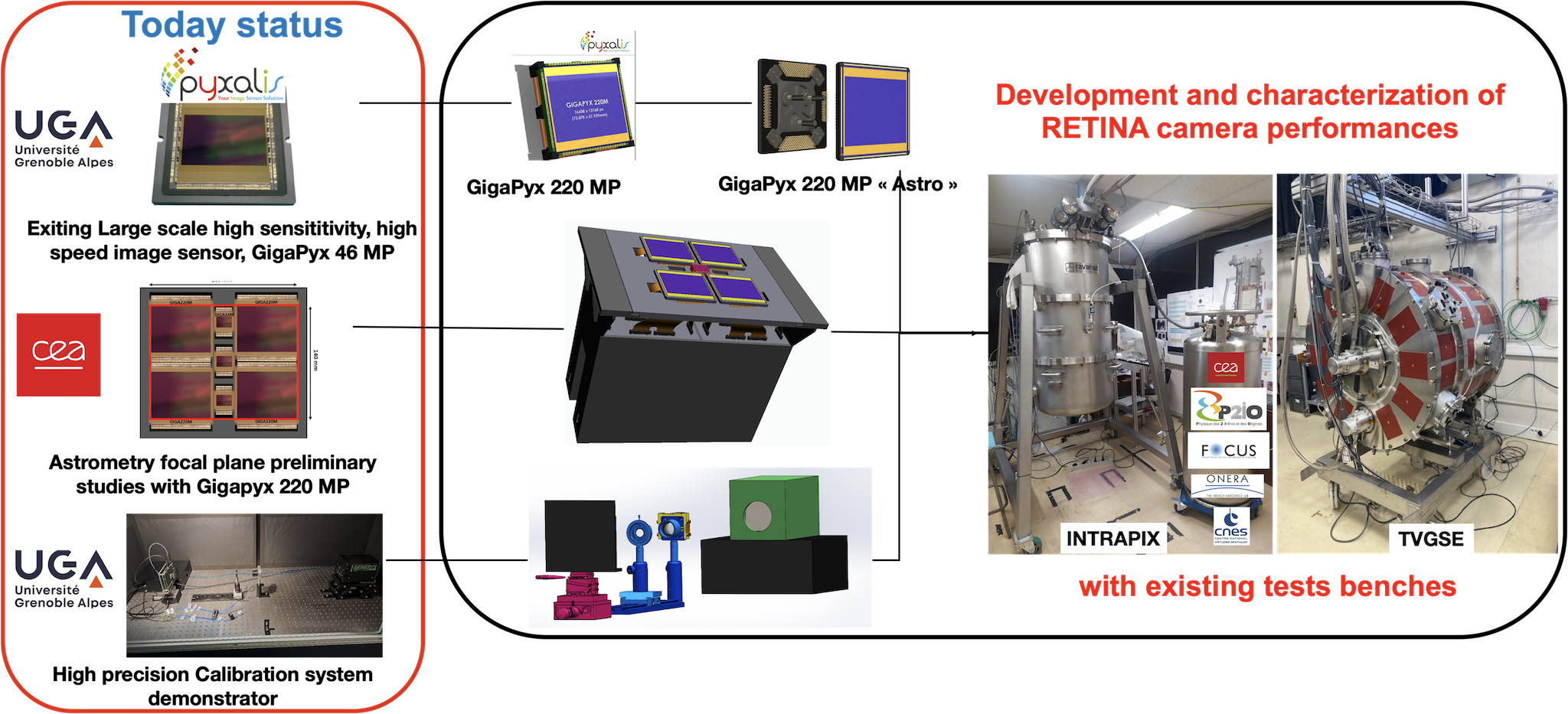}
  \caption{RETINA project: a collaborative initiative between IPAG, CEA-Saclay and Pyxalis to advance the Technology Readiness Level (TRL) of critical components for future astrometric missions funded by the PEPR Origins/ANR.}
  \label{fig:retina}
\end{figure}

At the heart of the effort to achieve sub-$\mu$as astrometry lies the RETINA project  (Fig.~\ref{fig:retina}), a collaborative initiative between IPAG, CEA-Saclay and Pyxalis to advance the Technology Readiness Level (TRL) of critical components for future astrometric missions. Central to RETINA is the development and validation of Pyxalis GIGAPYX CMOS detectors\cite{10.1117/12.3075196}, a family of large-format sensors specifically designed for high-precision astrometry.

The GIGAPYX 220\,MP detector, with its 220 millions pixel array ($16,640 \times 13,156$ pixels) and 4.4 $\mu$m pixel pitch, represents a key technological milestone. These detectors are intended as the building blocks for the gigapixel focal planes of Theia and HWO (instrument which we call HWO-AGATE for \emph{``Accurate Guiding and Astrometry Targeting ExoEarth'')}, where multiple sensors are tiled to cover the required field of view. The absence of microlenses in the GIGAPYX design is particularly critical, as it avoids the systematic distortions that microlenses can introduce into the intra-pixel response, which would otherwise compromise centroiding precision at the sub-$\mu$as level.

A comprehensive characterization campaign\cite{Lizzana2026_Calibrations} has been undertaken to validate the suitability of these detectors for astrometric applications (Sect.~\ref{sec:characterization}). Independent measurements performed at IPAG on the GIGAPYX-4600, the first member of the GIGAPYX family, have confirmed that the sensors meet or exceed the stringent performance requirements. The detectors exhibit readout noise as low as 12 electrons, linearity errors below 1\%, and pixel response non-uniformity under 1.3\%, all of which are essential for achieving the required precision. The dark current, while slightly higher than the manufacturer's specification at 36 $\pm$ 20 e$^{-}$/s at 35$^o$C, remains within acceptable limits and is expected to improve in future revisions. Most notably, the defective pixel rate is exceptionally low at 0.0004\%, minimizing the impact of bad pixels on scientific observations. The inter-pixel capacitance, measured at less than 3\%, introduces a manageable level of crosstalk that can be corrected through post-processing algorithms.

These results establish the GIGAPYX detectors as a robust technological foundation for the gigapixel focal planes of Theia and HWO-AGATE, capable of supporting the demanding requirements of sub-$\mu$as astrometry.

\section{MICROPIXEL-SCALE DETECTOR CHARACTERIZATION}
\label{sec:characterization}

Achieving the precision necessary to detect Earth-mass exoplanets requires more than just high-performance detectors; it demands an intimate understanding of detector behavior at the micropixel scale. Traditional assumptions about pixel geometry and response must be revisited, as even sub-pixel distortions can bias centroid measurements at the required precision level.

\begin{figure}[p]
  \centering
    \includegraphics[width=0.9\hsize]{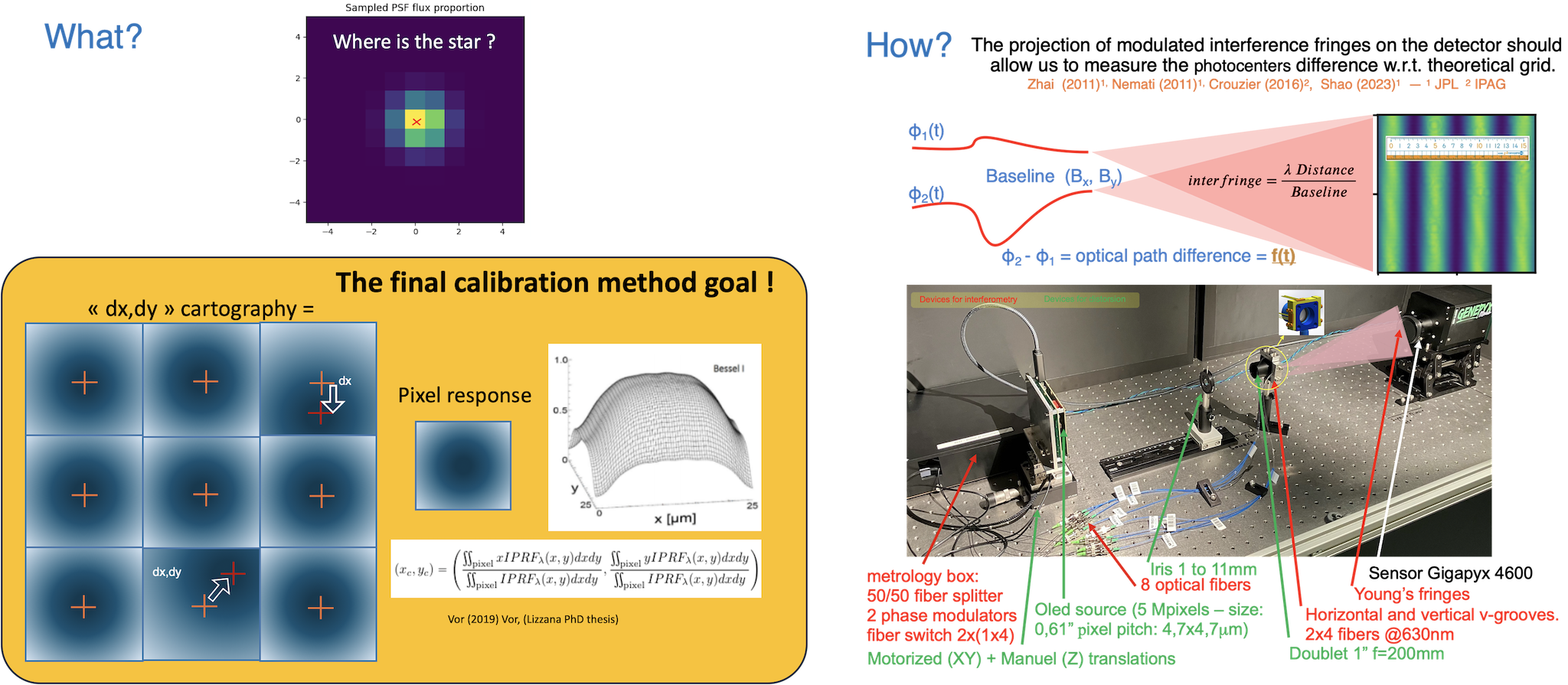}
  \caption{Left: at the pixel level, a map of pixel centroid displacements, correcting for geometric imperfections in the detector is required. Right: Young’s fringes method provides a map of pixel centroid displacements, correcting for geometric imperfections in the detector, and, a dedicated distortion bench has been developed, which projects a simulated star field onto the detector using a high-resolution micro-OLED screen.}
  \label{fig:detcalib}
\end{figure}

One of the most promising approaches to monitor pixel geometry and intra-pixel response across gigapixel-scale focal planes is the use of Young's fringe projection\cite{2011SPIE.8151E..0WN, 2016A&A...595A.108C, Shao2023_Microarcsecond}, an interferometric technique that enables the precise mapping of pixel centroid displacements\cite{Rousset2026_YoungsFringes} (Fig.~\ref{fig:detcalib}). By projecting a controlled interference pattern onto the detector and scanning it across the focal plane, this method allows for the measurement of intra-pixel response variations with extraordinary accuracy. Laboratory validations have demonstrated that this technique can achieve a precision of 5$\times$10$^{-4}$ pixel with the current setup, and ongoing work aims to push this down to the 5$\times$10$^{-5}$ pixel level (corresponding to $\sim 200$\,pm for 4.4 $\mu$m pixels), which is the requirement for HWO's astrometric instrument AGATE.

\begin{figure}[p]
  \centering \includegraphics[width=0.9\hsize]{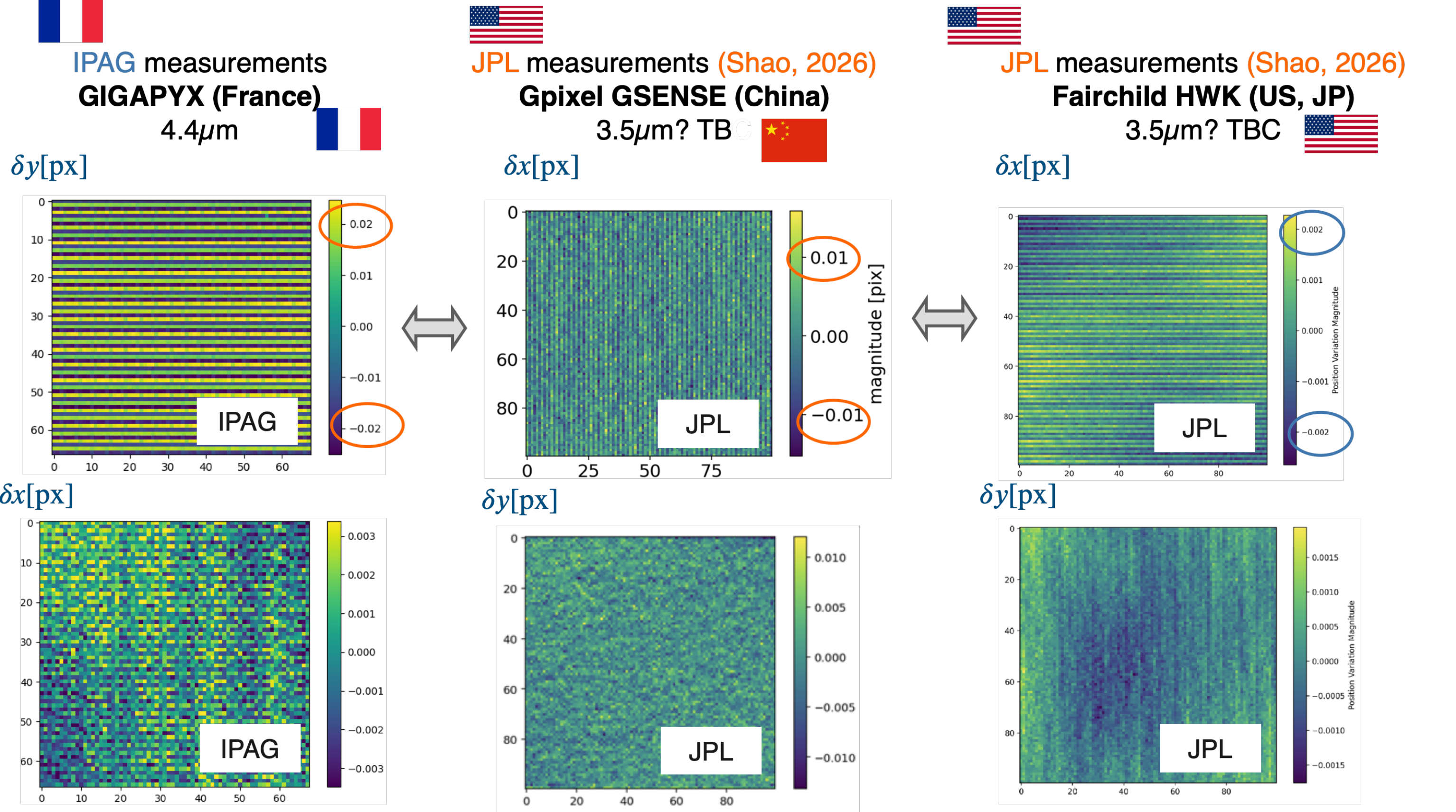} 
  \caption{Results from IPAG merasurements for the GIGAPYX 4600 detector in a $70\times70$ window compared to the tests published by Shao and colleagues\cite{Shao2026} for the Gpixel GSENSE detector and the Fairchild HWK detector.}
  \label{fig:calibcomp}
\end{figure}

The interferometric calibration bench developed at IPAG employs a HeNe laser whose light is split into two fibers, with the phase difference between the fibers modulated to scan the interference fringes across the detector. By analyzing the temporal response of each pixel to the moving fringes, the pixel centroid displacements ($\delta x$, $\delta y$) can be derived. This method is not only highly precise but also compatible with onboard implementation, making it a prime candidate for in-flight calibration during mission operations. Figure \ref{fig:calibcomp} shows the results obtained on our IPAG bench for the GIGAPYX 4600 detector when using a $70\times70$ window. These results are compared to the tests published by Shao et al.\cite{Shao2026} for similar windows on Gpixel GSENSE and Fairchild HWK detectors, which have comparable pixel pitches. Repeatability measurements show that we achieved $6.10^{-4}$\,px precision in the horizontal direction of the GIGAPYX 46MP detector, similar to that obtained by Shao et al. with their detectors. In the vertical direction, we measured a precision of $1.10^{-3}$\,px; however, we believe that our results are still limited by the inadequately calibrated inter-pixel capacitance\cite{Rousset2026_YoungsFringes}. 

A key challenge in scaling this technique to gigapixel focal planes is the paraxial approximation, which assumes straight fringes across the detector. For large detectors, the fringes are inherently hyperbolic, and the approximation breaks down. To address this, the RETINA team is developing a hyperbolic fringe model that accounts for the true geometry of the interference pattern, ensuring accuracy across the entire focal plane\cite{Rousset2026_YoungsFringes}.

\section{CALIBRATING OPTICAL DISTORTION AND REFINING DIFFERENTIAL ASTROMETRY}
\label{sec:calibration}

While detector characterization is critical, it represents only one piece of the puzzle. To convert pixel-level measurements into angular positions on the sky, the optical distortion introduced by the telescope and instrument must be precisely calibrated. This is particularly challenging for large-field instruments, where distortions can vary significantly across the field of view.

We employ a multi-level calibration strategy to address this. At the pixel level, the Young's fringes method provides a map of pixel centroid displacements, correcting for geometric imperfections in the detector. At the telescope level, optical distortion is calibrated using the reference stars as metrology sources\cite{Shao2023_Microarcsecond}. To demonstrate the calibration process for the optics, a dedicated distortion bench has been developed, which projects a simulated star field onto the detector from a high-resolution micro-OLED screen. By introducing controlled distortions via a mobile translating diaphragm, the bench allows for the validation of polynomial models that describe the optical distortion across the field of view.

To refine differential astrometry methods, our team has developed a global astro-calibration framework\cite{Lizzana2026_Calibrations} that jointly estimates stellar positions, telescope attitude, plate scale, and optical distortion. This framework treats the calibration problem as a unified inverse problem, where all parameters are optimized simultaneously to minimize the residuals between observed and modeled star positions. By leveraging the stars in the field of view as metrology sources, similar to the approach used by the Gaia mission\cite{2012A&A...538A..78L, 2023A&A...674A...1G}, this method ensures that systematic errors are controlled at the sub-$\mu$as level.

\begin{figure}[t]
  \centering
  \includegraphics[width=0.9\hsize]{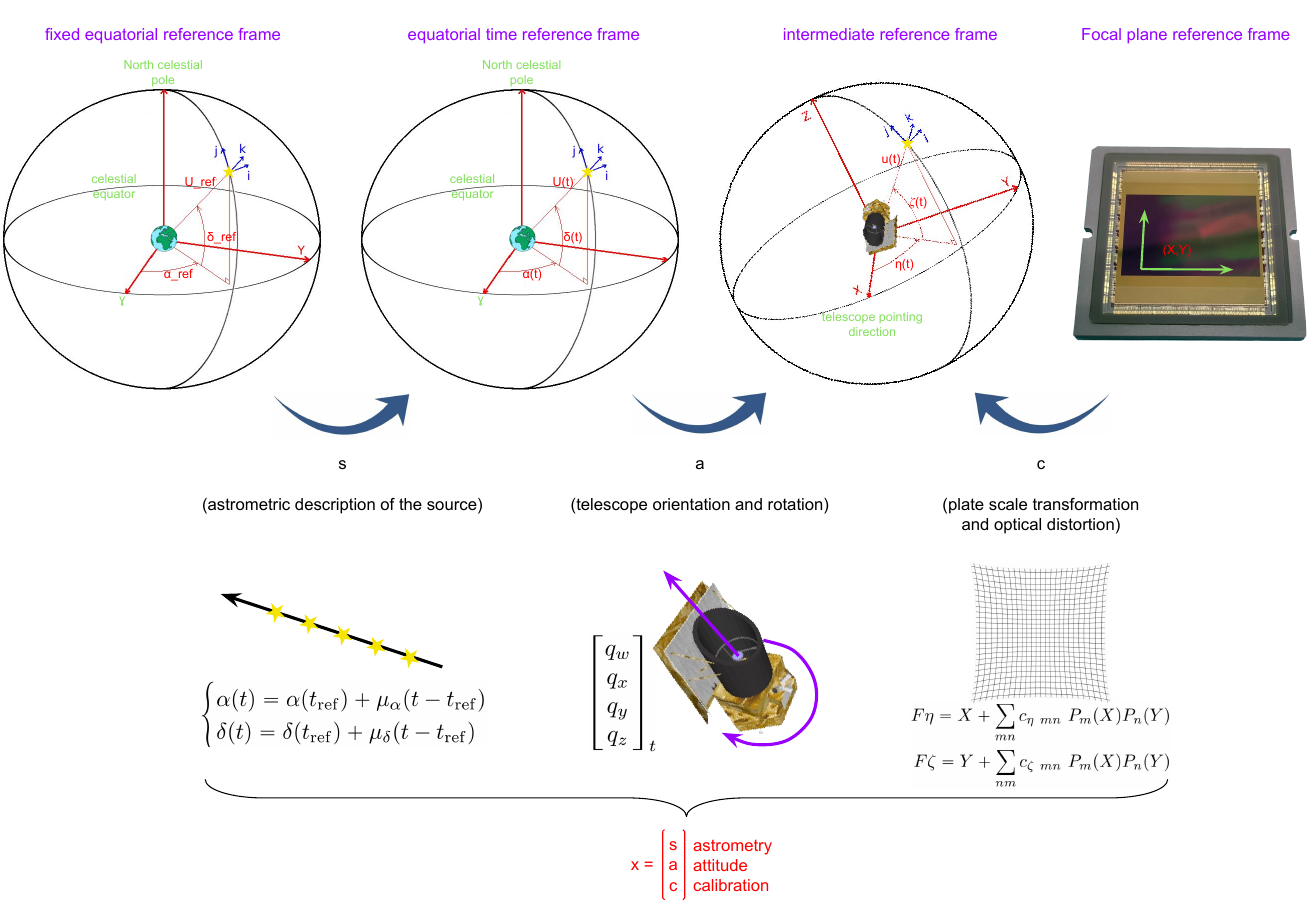}
  \caption{Illustration of the astro-calibration method\cite{Lizzana2026_Calibrations}. The four reference frames are shown. The three models ($s$ the stellar reference sources, $a$ the telescope attitude, and $c$ the telescope distortion) that link them and constitute the parameter vector $x$ are also represented.}
  \label{fig:astrocalib}
\end{figure}

The framework operates within a hierarchy of reference systems, connecting the equatorial reference frame (star positions at a reference epoch) to the detector frame (pixel coordinates) through a series of transformations. These transformations account for stellar proper motions, telescope attitude, and instrumental distortions, and are solved iteratively using a combination of nonlinear least squares and weighted linear regression (Fig.~\ref{fig:astrocalib}). Preliminary laboratory tests have demonstrated the feasibility of this approach, and ongoing simulations are refining its performance.

\section{SYSTEM-LEVEL INTEGRATION AND LABORATORY VALIDATIONS}
\label{sec:integration}

The ultimate goal is to integrate these advancements into a coherent system capable of meeting the sub-$\mu$as precision requirements of Theia and HWO-AGATE\cite{ArdellierDesages2026_TechMaturation}. This involves not only validating individual components but also ensuring that they work together seamlessly at the system level.

To this end, a suite of characterization benches has been developed to support the validation of detector performance and calibration methods\cite{Pancher2026_Benches}. The flat-field calibration bench uses an integrating sphere to provide uniform illumination for electro-optical characterization, while the interferometric bench (Sect.~\ref{sec:characterization}) enables the precise measurement of pixel centroid displacements. A distortion bench (Sect.~\ref{sec:calibration}) complements these by validating optical distortion models using simulated star fields. These benches are designed to be modular and scalable, with upgrades underway to support multi-detector configurations representative of the gigapixel focal planes of Theia and HWO-AGATE.

A critical aspect of system-level integration is the error budget\cite{Amiaux2026_SystemAnalysis}, which allocates the total allowable error to individual contributors. For HWO-AGATE, the single-measurement error must be $\leq$0.5 $\mu$as to achieve a signal-to-noise ratio of at least 6 after 100 revisits. This budget is divided among photocenter determination errors (for both the target and calibration stars) and calibration errors (including pixel loaction and optical distortion corrections). The RETINA project is working to ensure that each of these contributors is controlled at the sub-$\mu$as level, with a particular focus on systematic errors, which must be mitigated through precise calibration.

The instrument architecture for HWO's astrometric mode AGATE reflects these requirements. The focal plane will consist of four 220 MP detectors surrounding a central 14 MP detector, providing a total of \~1 gigapixel across a 3 $\times$ 4 arcmin$^2$ field of view. The central detector, operating in a fast windowed readout mode, will measure the bright target star, while the surrounding detectors, in full-frame mode, will track the fainter calibration stars. This configuration ensures that the instrument can achieve the necessary dynamic range and precision to detect Earth-mass exoplanets.

% \section{SYNERGY BETWEEN DETECTOR TECHNOLOGY, OPTICAL CALIBRATION, AND MISSION DESIGN}
% \label{sec:synergy}

One of the RETINA project's defining features is its comprehensive approach to astrometric instrument development. Rather than treating detector technology, optical calibration, and mission design as separate endeavors, the project emphasizes their interconnectedness, recognizing that advancements in one area can facilitate progress in the others. This synergy extends to the mission design as well. The field of view, pixel scale, and readout strategies of Theia and HWO-AGATE have been optimized based on the performance of the GIGAPYX detectors and the capabilities of the calibration methods. For instance, the 3 $\times$ 4 arcmin$^2$ field of view for HWO-AGATE was selected to maximize the number of reference stars available for differential astrometry, while the 6.2\,mas/pixel scale ensures that the point spread function is adequately sampled\cite{Amiaux2026_SystemAnalysis}. The fast readout mode for the central detector and the full-frame mode for the surrounding detectors are designed to accommodate the wide range of stellar magnitudes encountered in astrometric observations, from bright target stars (magnitude 3--7) to faint calibration stars (magnitude 15--20).

\section{CONCLUSION AND PERSPECTIVES}
\label{sec:conclusion}

The RETINA project represents a significant leap forward in the development of high-precision astrometric instruments for future space missions. By advancing the TRL of GIGAPYX CMOS detectors\cite{ArdellierDesages2026_TechMaturation,Lizzana2026_Calibrations}, developing novel calibration methods such as Young's fringe projection for large format detector\cite{Rousset2026_YoungsFringes}, and refining differential astrometry techniques\cite{Amiaux2026_SystemAnalysis, Pancher2026_Benches}, the project is paving the way for the sub-$\mu$as precision required to detect Earth-mass exoplanets and probe dark matter dynamics.

The technical strategies employed by RETINA, ranging from micropixel-scale detector characterization to system-level integration, are designed to address the most pressing challenges in high-precision astrometry. Laboratory validations have already demonstrated the feasibility of many of these approaches, and ongoing work is focused on scaling them to the gigapixel focal planes of Theia and HWO-AGATE.

Looking ahead, the next critical steps include the validation of calibration methods on 220\, MP detectors, the development of multi-detector configurations, and the achievement of TRL 5 for all key technologies by the HWO Mission Consolidation Review in 2029. By enabling the detection and characterization of Earth-like exoplanets, RETINA will contribute to a transformative understanding of exoplanetary systems and the role of dark matter in galactic dynamics.

Ultimately, the success of the RETINA project underlines the importance of synergy in instrument development. By integrating advancements in detector technology, optical calibration, and mission design, the project demonstrates how a comprehensive approach can overcome the formidable challenges of sub-$\mu$as astrometry and unlock new frontiers in astrophysics.

\acknowledgments

The authors gratefully acknowledge the contributions of the RETINA team of researchers and engineers for their invaluable support and collaboration throughout this work. This collaborative effort is supported by the CEA, CNRS, and Université Grenoble Alpes (UGA). The technology maturation program RETINA receives support from the French Government through the France 2030 investment plan, operated by the French National Research Agency under grant ANR-25-EXOR-0003. The work has been also partially supported by the LabEx FOCUS by the French National Research Agency through the grant ANR-11-LABX-0013 and in the framework of the "France 2030” program by the grant ANR-15-IDEX-02. The authors acknowledge financial support from the Centre National d’\'Etudes Spatiales in France (CNES, ROR: \url{https://ror.org/04h1h0y33}), thanks to the Research and Technology (R\&T) and mission preparation programs.  

\bibliography{malbet_icso2026}
\bibliographystyle{spiebib}

\end{document}